\documentclass[12pt]{article} 

\baselineskip=\normalbaselineskip

\usepackage{mathrsfs,amsbsy,amssymb,braket,bm,latexsym,amsfonts,amsmath,amsthm}
\usepackage{graphicx,color}

\makeatletter
\catcode`\@=11
\@addtoreset{equation}{section}

\def\@seccntformat#1{\csname the#1\endcsname.~~}
\makeatother

\begin{document}
\begin{titlepage}
\renewcommand{\thefootnote}{\fnsymbol{footnote}}
\begin{flushright}
KUNS-2725
\end{flushright}
\vspace*{1.0cm}

\begin{center}
{\Large \bf 
Gradient flow and the renormalization group
}
\vspace{1.0cm}

\centerline{
{Yoshihiko Abe}%
\footnote{E-mail address: 
y.abe@gauge.scphys.kyoto-u.ac.jp} and
{Masafumi Fukuma}%
\footnote{E-mail address: 
fukuma@gauge.scphys.kyoto-u.ac.jp}%
}

\vskip 0.8cm
{\it Department of Physics, Kyoto University, Kyoto 606-8502, Japan}
\vskip 1.2cm

\end{center}

\begin{abstract}

We investigate the renormalization group (RG) structure 
of the gradient flow. 
Instead of using the original bare action to generate the flow,  
we propose to use the effective action at each flow time. 
We write down the basic equation for scalar field theory 
that determines the evolution of the action, 
and argue that the equation can be regarded as a RG equation 
if one makes a field-variable transformation at every step 
such that the kinetic term is kept to take the canonical form. 
We consider a local potential approximation (LPA) to our equation, 
and show that the result has a natural interpretation with Feynman diagrams. 
We make an $\varepsilon$ expansion of the LPA  
and show that it reproduces the eigenvalues of the linearized RG transformation 
around both the Gaussian and the Wilson-Fisher fixed points 
to the order of $\varepsilon$. 

\end{abstract}
\end{titlepage}

\pagestyle{empty}
\pagestyle{plain}

\tableofcontents
\setcounter{footnote}{0}

\section{Introduction}
\label{sec:introduction}

In recent years the gradient flow has attracted much attention 
for practical and conceptual reasons 
\cite{Luscher:2009eq,Luscher:2010iy,Luscher:2011bx,Luscher:2013cpa,
Suzuki:2013gza,Makino:2014taa,Kikuchi:2014rla}.
Practically, as shown by L\"uscher and Weisz 
\cite{Luscher:2010iy,Luscher:2011bx}, 
the gradient flow in nonabelian gauge theory 
does not induce extra UV divergences in the bulk, 
so that the bulk theory is finite 
once the boundary theory is properly renormalized. 
Hence the ultralocal products of bulk operators 
automatically give renormalized composite operators, 
and this fact yields a lot of applications 
including a construction of energy-momentum tensor on the lattice 
\cite{Suzuki:2013gza,Makino:2014taa}.

On the other hand, there has been an expectation 
that the gradient flow may be interpreted 
as a renormalization group (RG) flow 
(see, e.g., \cite{Luscher:2013vga,Kagimura:2015via,Yamamura:2015kva,
Aoki:2016ohw,Makino:2018rys}). 
This expectation is based on the observation 
made in \cite{Luscher:2010iy}. 
To see this, 
let us consider a Euclidean scalar field theory in $d$ dimensions 
with the bare action $S_0[\phi]$. 
We assume that the theory is implemented with some UV cutoff $\Lambda_0$. 
The gradient flow is then given by 
\begin{align}
 \partial_\tau\phi_\tau(x) &={} -\frac{\delta S_0}{\delta \phi(x)}[\phi_\tau],
 \quad
 \phi_{\tau=0}(x) = \phi_0(x).
\label{flow0}
\end{align}
If the field is canonically normalized as 
$\int_x [(1/2)(\partial_\mu\phi)^2+\cdots]$, 
then the flow equation gives a heat equation with perturbation: 
\begin{align}
 \partial_\tau\phi_\tau(x) &= \partial_\mu^2\phi_\tau(x) +~\cdots,
\label{heat_eq}
\end{align}
which can be solved as%
\footnote{
 In this paper we only consider scalar field theory, 
 but our discussion should be easily extended to other field theories. 
 We use a standard polymorphic notation; 
 $\int_x$ represents $\int d^d x$ when $x$ are spacetime coordinates 
 while $\int_p$ stands for $\int d^d p/(2\pi)^d$ when $p$ are momenta. 
 We often denote $\phi(x)$ by $\phi_x$.
} 

\begin{align}
 \phi_\tau(x)
 = \int_y K_\tau(x-y)\,\phi_0(y)
 + \cdots, 
\end{align}
where $K_\tau(x-y)$ is the heat kernel: 
\begin{align}
 K_\tau(x-y)  = \int_p e^{i p(x-y) - \tau\,p^2}
 = \frac{1}{(4\pi\tau)^{d/2}}\,e^{-(x-y)^2/4\pi\tau}.
\end{align}
Thus, $\phi_\tau(x)$ can be interpreted as an effective field 
which is coarse-grained from $\phi_0(y)$ within the radius $r\propto \sqrt{\tau}$.

However, this interpretation is not perfectly matched 
with the philosophy of the renormalization group. 
In fact, if we denote the solution to \eqref{flow0} 
by $\phi_\tau(\phi_0)=\bigl(\phi_\tau(x;\phi_0)\bigr)$ 
so as to specify its initial value, 
the distribution function of $\phi$ at time $\tau$ will be given by 
\begin{align}
 p_\tau[\phi] = \frac{1}{Z_0}\,\int [d\phi_0]\,
 \delta[\phi-\phi_\tau(\phi_0)]\,
 e^{-S_0[\phi_0]}
 \quad\Bigl( Z_0 \equiv \int [d\phi_0]\, e^{-S_0[\phi_0]}\Bigr).
\end{align}
The flow equation gives the field $\phi$ a tendency 
to approach the classical solution of the original bare action $S_0[\phi]$, 
and thus $p_\tau[\phi]$ will take a sharp, $\delta$ function-like peak 
at the classical solution in the large $\tau$ limit, 
but this is not what we expect in the renormalization group; 
$\phi_\tau$ at large $\tau$ should be regarded as a low-energy effective field, 
which can be well treated as the classical solution 
to the low-energy effective action at scale $\Lambda=1/\sqrt{\tau}$, 
not to the bare action which itself can be regarded 
as giving an effective theory at the original cutoff $\Lambda_0\,(\gg\Lambda)$.

In this paper, 
we propose a novel gradient flow 
that gives the field a tendency to approach 
the classical solution of the effective action 
at scale $\Lambda=1/\sqrt{\tau}$ when the derivative is taken: 
\begin{align}
 \partial_\tau\phi_\tau(x) &={} -\frac{\delta S_\tau}{\delta \phi(x)}[\phi_\tau],
 \quad
 \phi_{\tau=0}(x) = \phi_0(x).
\label{flow1}
\end{align}
Assuming that the initial value $\phi_0(x)$ is distributed 
according to the distribution function $e^{-S_0[\phi_0]}/Z_0$, 
we impose the self-consistency condition 
that the classical solution $\phi_\tau(x)$ be distributed 
with $e^{-S_\tau[\phi]}/Z_\tau$:%
\footnote{ 
 Note that the partition function is constant in time, 
 $Z_\tau \equiv \int[d\phi]\,e^{-S_\tau[\phi]} = Z_0$.
} 
\begin{align}
 e^{-S_\tau[\phi]} \equiv \int [d\phi_0]\,\delta[\phi-\phi_\tau(\phi_0)]\,
 e^{-S_0[\phi_0]}, 
\label{consistency0a}
\end{align}
where $\phi(x)$ should have only the coarse-grained degrees of freedom. 
We investigate the consequences of this requirement, 
and argue that the obtained equation for $S_\tau[\phi]$ 
may be regarded as a RG equation 
if one makes a field-variable transformation at every step 
such that the kinetic term is kept to take the canonical form.

This paper is organized as follows. 
In Section \ref{sec:formulation} 
we write down the basic equation that determines 
the evolution of $S_\tau[\phi]$. 
In Section \ref{sec:LPA} 
we consider a local potential approximation (LPA) to our equation, 
and show that the result has a nice interpretation with Feynman diagrams. 
In Section \ref{sec:epsilon} 
we make an $\varepsilon$ expansion of the LPA 
and show that it reproduces the eigenvalues of the linearized RG transformation 
around both the Gaussian and the Wilson-Fisher fixed points 
to the order of $\epsilon$. 
Section \ref{sec:conclusion} is devoted to conclusion and outlook.

\section{Formulation}
\label{sec:formulation}

We first rewrite the consistency condition \eqref{consistency0a} 
to a differential form:%
\footnote{ 
 In this paper, 
 in order to simplify discussions, 
 we do not seriously take into account the anomalous dimension $\gamma=\eta/2$, 
 which may be incorporated by adding a term 
 $(\gamma/2\tau)\,\phi_\tau(x)$ 
 to the right-hand side of the first equation in \eqref{flow1}.
} 
\begin{align}
 \partial_\tau e^{-S_\tau[\phi_\tau]}
 &=\int [d\phi_0] \int_x \Bigl(\frac{\delta}{\delta \phi(x)}
 \delta\bigl[\phi-\phi_\tau(\phi_0)\bigr]\Bigr)\,
 \bigl( -\partial_\tau \phi_\tau(x)\bigr)e^{-S_0[\phi_0]}
\nonumber
\\
 &=\int [d\phi_0]\int_x 
 \Bigl(\frac{\delta}{\delta \phi(x)} \delta 
 \bigl[ \phi-\phi_\tau\left( \phi_0\right)\bigr]\Bigr)\,
 \frac{\delta S_\tau}{\delta \phi(x)} [\phi_\tau ]\,
 e^{-S_0[\phi_0]}
\nonumber
\\
 &=\int_x \frac{\delta}{\delta \phi(x)}
 \Bigl[ \frac{\delta S_\tau[\phi]}{\delta \phi(x)}\,e^{-S_\tau[\phi]}\Bigr],
\label{flow1a}
\end{align}
which in turn gives the following differential equation for $S_\tau[\phi]$: 
\begin{align}
 \partial_\tau S_\tau [\phi]
 &=
 \int_x
 \Bigl[{} - \frac{\delta^2 S_\tau[\phi]}{\delta \phi(x)^2}
 + \frac{\delta S_\tau[\phi]}{\delta \phi(x)}
 \frac{\delta S_\tau[\phi]}{\delta \phi(x)}\Bigr].
\label{flow1b}
\end{align}
However, one can easily see that UV divergences arise 
from the second-order functional derivative at the same point, 
$\delta^2 S/\delta\phi(x)^2$. 
The reason why such UV divergences appear in the effective theory 
is that we have not taken into account the fact 
that $\phi(x)$ should have only the coarse-grained degrees of freedom 
with cutoff $\Lambda=1\/\sqrt{\tau}$.

To see how to incorporate this fact, 
it is helpful to consider a sharp cutoff for a while, 
instead of the smooth smearing with the heat kernel $K_\tau(x-y)$. 
Namely, we assume that the flowed field is cut off as 
$\phi_\tau(x)=\int_{|p|\leq 1/\sqrt{\tau}} e^{i p x}\,\phi_{\tau,p}$, 
and accordingly that the action $S_\tau[\phi]$ depends 
only on the lower modes $\phi_p$ $(|p|\leq 1/\sqrt{\tau})$ 
of the scalar field $\phi(x)=\int_p e^{i p x}\,\phi_{p}$.  
Then, the calculation in \eqref{flow1a} will be modified as 
\begin{align}
 \partial_\tau e^{-S_\tau[\phi]}
 &=\int [d\phi_0] \int_{|p|\leq 1/\sqrt{\tau}} 
 \Bigl(\frac{\delta}{\delta \phi_p}
 \delta[\phi-\phi_\tau(\phi_0)] \Bigr)\,
 \bigl( -\partial_\tau \phi_{\tau,p}\bigr)e^{-S_0[\phi_0]}
\nonumber
\\
 &=\int [d\phi_0] \int_{|p|\leq 1/\sqrt{\tau}} 
 \Bigl(\frac{\delta}{\delta \phi_p}
 \delta[\phi-\phi_\tau(\phi_0)] \Bigr)\,
 \frac{\delta S_\tau}{\delta \phi_{-p}} [\phi_\tau ]\,
 e^{-S_0[\phi_0]}
\nonumber
\\
 &=\int_{|p\leq 1/\sqrt{\tau}} \frac{\delta}{\delta \phi_p}
 \Bigl[ \frac{\delta S_\tau[\phi]}{\delta \phi_{-p}}\,e^{-S_\tau[\phi]}\Bigr].
\label{flow2a_sharp}
\end{align}
Returning back to the smooth cutoff with the heat kernel, 
eq.~\eqref{flow2a_sharp} will be expressed as 
\begin{align}
 \partial_\tau e^{-S_\tau[\phi]}
 &=\int_{x,y}\ K_\tau(x-y)\,\frac{\delta}{\delta \phi(x)}
 \Bigl[\frac{\delta S_\tau[\phi]}{\delta \phi(y)}\,
 e^{-S_\tau[\phi]}\Bigr],
\label{flow2a}
\end{align}
which is equivalent to the equation
\begin{align}
 \partial_\tau S_\tau[\phi]
 &=\int_{x,y} K_\tau(x-y)\Bigl[
 \frac{\delta S_\tau[\phi]}{\delta \phi(x)}
 \frac{\delta S_\tau[\phi]}{\delta \phi(y)}
 -\frac{\delta^2 S_\tau[\phi]}{\delta \phi(x)\delta \phi(y)}\Bigr].
\label{flow2b}
\end{align}
We see that there no longer exist divergences of the aforementioned type.
For the rest of this paper, 
we treat \eqref{flow2b} as the equation 
that {\em defines} the flow of $S_\tau(\phi)$.

We here make an important comment that \eqref{flow2a} can be rewritten 
in the form of Fokker-Planck equation: 
\begin{align}
 \partial_\tau e^{-S_\tau [\phi]}
 &=\int_{x,y}\,K_\tau(x-y)\,
 \Bigl[ \frac{\delta^2S[\phi]}{\delta \phi(x)\delta \phi(y)}
 -\frac{\delta S_\tau[\phi]}{\delta \phi(x)}
 \frac{\delta S_\tau[\phi]}{\delta \phi(y)}\Bigr]\,e^{-S_\tau[\phi]}
\nonumber
\\
 &=\int_{x,y}\,\frac{\delta}{\delta \phi(x)}\,K_\tau(x-y)\,
 \Bigl[ \frac{\delta}{\delta \phi(y)}
 +2\frac{\delta S_\tau[\phi]}{\delta \phi(y)}\Bigr]\, e^{-S_\tau[\phi]},
\label{FK1}
\end{align}
which corresponds to the Langevin equation 
\begin{align}
 \partial_\tau \phi_\tau(x)
 &= \nu_\tau(x) - 2\,\int_y K_\tau(x-y)
 \frac{\delta S_\tau[\phi]}{\delta \phi(y)}  
\label{Langevin0}
\end{align}
with the Gaussian white noise $\nu_\tau(x)$ normalized as 
\begin{align}
 \Braket{\nu_\tau(x)\nu_{\tau'}(y)}_\nu
 &= 2\, \delta( \tau-\tau')\,K_\tau(x-y).
\end{align}
The solution $\phi_\tau(x)$ to the Langevin equation 
now depends on the noise $\nu_\tau(x)$ 
as well as the initial value $\phi_0(x)$, 
\begin{align}
 \phi_\tau(x)=\phi_\tau\left(x;\phi_0,\nu\right).
\end{align}
Then, denoting the Gaussian measure of $\nu$ by $[d\rho(\nu)]$, 
the distribution function $e^{-S_\tau[\phi]}/Z_\tau$ 
[see \eqref{consistency0a}] 
can also be written as 
\begin{align}
 e^{-S_\tau[\phi]} &= \int [d\phi_0]\,
 \bigl\langle \delta[\phi-\phi_\tau(\phi_0,\nu)] \bigr\rangle_\nu\,
 e^{-S_0[\phi_0]}
\nonumber
\\
 &= \int [d\phi_0][d\rho(\nu)]\,
 \delta[\phi-\phi_\tau(\phi_0,\nu)] \,
 e^{-S_0[\phi_0]}.
\label{consistency0b}
\end{align}
The Langevin equation \eqref{Langevin0} shows that 
the field $\phi_\tau(x)$ makes a random walk due to the noise term, 
but at the same time it tries to approach 
the classical solution to $S_\tau[\phi]$. 
We thus find the mathematical equivalence 
between two expressions \eqref{consistency0a} and \eqref{consistency0b} 
that have different meanings; 
the former is purely deterministic in the course of evolution 
while the latter is stochastic. 
This observation may support an idea 
that a seemingly deterministic evolution is actually accompanied 
by an integration over some fluctuating degrees of freedom.

\section{Local potential approximation}
\label{sec:LPA}

In order to investigate how our equation \eqref{flow2b} works as a RG equation, 
we make a local potential approximation 
\cite{Wilson:1973jj,Hasenfratz:1985dm,Morris:1994ie}:
\begin{align}
 S_\tau[\phi]&=
 \int_x \Bigl[ V_\tau(\phi_x) + \frac{1}{2}\, ( \partial_\mu \phi_x)^2 \Bigr].
\label{LPA0}
\end{align}
The canonical form of kinetic term is particularly important for our purpose 
to interpret the gradient flow as a RG flow 
[see discussions around \eqref{heat_eq}]. 
However, even when we normalize the field $\phi_x$ in this way at time $\tau$, 
the action may no longer take a canonical form at $\tau+\epsilon$. 
In order for the interpretation $\Lambda=1/\sqrt{\tau}$ to hold 
also at time $\tau+\epsilon$ 
[i.e.\ $\Lambda-\delta\Lambda=1/\sqrt{\tau+\epsilon}
= (\tau\,e^{\epsilon/\tau})^{-1/2}$], 
we then need to make a field-variable transformation at $\tau+\epsilon$
to retain the kinetic term in the canonical form.

For making necessary calculations, 
it is convenient to start from the local potential approximation 
of the second order: 
\begin{align}
 I_\tau[\varphi] &\equiv
 \int_x \Bigl[ U_\tau(\varphi_x)
 + \frac{1}{2}\, W_\tau(\varphi_x)\,( \partial_\mu \varphi_x)^2 \Bigr]
\label{LPA2}
\end{align}
and to investigate the evolution of 
$U_\tau(\varphi)$ and $W_\tau(\varphi)$ 
from $\tau$ to $\tau+\epsilon$ 
with the initial values 
$U_\tau(\varphi)=V_\tau(\varphi)$ and $W_\tau(\varphi)=1$. 
One can easily derive the following combined equations:%
\footnote{ 
 Among formulas that may be useful in deriving the equations are
 \begin{align}
  &\partial_x^2 K_\tau (x-y) = \partial_\tau K_\tau(x-y),
  \qquad
  \int_{x-y} K_\tau(x-y)\,(x-y)_\mu (x-y)_\nu = 2\,\tau\,\delta_{\mu\nu},
 \nonumber
 \\
  &\int_{x,y} K_\tau (x-y)\,f(\phi_x)\,g(\phi_y)
  = \int_x\bigl[\, f(\phi_x)\,g(\phi_x)
  - \tau\,(\partial_\mu\phi_x)^2\,f'(\phi_x)\,g'(\phi_x)
  + O(\tau^2) \bigr].
  \nonumber
 \end{align}
} 
\begin{align}
 \partial_\tau U_\tau(\varphi)&
 = U_\tau'(\varphi)^2
 -\frac{1}{(4\pi\tau)^{d/2}}\,U''_\tau(\varphi)
 -\frac{d}{2\tau}\,\frac{1}{(4\pi\tau)^{d/2}}\,W(\varphi),
\label{RG1}
\\
 \partial_\tau W_\tau(\varphi)&
 =2\, U'_\tau(\varphi)\, W'_\tau(\varphi)
 +4\, U''_\tau(\varphi)\, W_\tau(\varphi)
 -2\, \tau\, U''_\tau(\varphi)^2
 -\frac{1}{(4\pi\tau)^{d/2}}\,W''_\tau(\varphi).
\label{RG2}
\end{align}
From these, 
we find that the coefficient of $(1/2) (\partial_\mu\varphi_x)^2$ 
changes from the normalized value $W_\tau(\varphi)\equiv 1$ 
to
\begin{align}
 W_{\tau+\epsilon}(\varphi)
 &= 1+\epsilon\,\partial_\tau W_\tau(\varphi)
 =1 + \epsilon\bigl[ 4\,U_\tau''(\varphi)
 - 2\tau\,U_\tau''(\varphi)^2\bigr]
\nonumber
\\
 &\equiv 1 + 2\epsilon\, \rho_\tau'(\varphi).
\end{align}
Thus, the canonically normalized field $\phi$ at $\tau+\epsilon$ 
is given by integrating the equation 
$d\phi/d\varphi=\sqrt{W_{\tau+\epsilon}(\varphi)}
=1 + \epsilon\,\rho_\tau'(\varphi)$, 
and we find the following relation to the order of $\epsilon$: 
\begin{align}
 \varphi &= \phi - \epsilon\,\rho_\tau(\phi) 
 = \phi - \epsilon\,\int_0^{\phi} d\phi\,
 \bigl[ 2\,U_\tau''(\phi) - \tau\,U_\tau''(\phi)^2\bigr].
\end{align}
The Jacobian%
\footnote{ 
 The prime means that the determinant or the trace 
 should be taken on the partial functional space 
 under the projection of $K_\tau(x-y)$. 
} 
${\rm Det}'\,(\delta\varphi/\delta\phi)
= e^{{\rm Tr}'\,\log\,(\delta\varphi/\delta\phi)}$ 
is calculated with 
\begin{align}
 {\rm Tr}'\,\log(\delta\varphi/\delta\phi)
 = \int_{x,y} K_\tau(x-y)\,\log\bigl[ 1-\epsilon\,\rho_\tau'(\phi_x)\bigr]
 \,\delta^d(x-y)
 ={}- \epsilon \int_x \frac{1}{(4\pi\tau)^{d/2}}\,\rho_\tau'(\phi_x).
\end{align}

By putting everything together, 
the change of the local potential for the canonically normalized field $\phi$ 
is given as follows 
[recall the initial condition $U_\tau(\phi)=V_\tau(\phi)$]:
\begin{align}
 V_{\tau+\epsilon}(\phi)
 &= \bigl[\,U_\tau(\varphi)
 + \epsilon\,\partial_\tau U_\tau(\varphi)\bigr]
 \bigr|_{\varphi=\phi-\epsilon\,\rho_\tau(\phi)}
 + \epsilon\,\frac{1}{(4\pi\tau)^{d/2}}\,\rho'_\tau(\phi)
\nonumber
\\
 &= V_\tau(\phi)
 + \epsilon\,\Bigl[
 - V_\tau'(\phi)^2 + \frac{1}{(4\pi\tau)^{d/2}}\,V_\tau''(\phi)
 + \tau\,V_\tau'(\phi)\,\int_0^\phi d\phi\,V_\tau''(\phi)^2
\nonumber
\\
 &~~~ - \frac{\tau}{(4\pi\tau)^{d/2}}\,V_\tau''(\phi)^2 
 - \frac{d}{2\,\tau(4\pi\tau)^{d/2}} \Bigr]. 
\label{dimful}
\end{align}
Note that the terms $V_\tau'(\phi)^2$ and $V_\tau''(\phi)$ 
appear in \eqref{dimful} 
as ${}-V_\tau'(\phi)^2 + {\rm const.}\,V_\tau''(\phi)$ 
that have the same signs as those in the Polchinski equation 
\cite{Polchinski:1983gv}, 
although the signs of the terms $U_\tau'(\phi)^2$ and $U_\tau''(\phi)$ 
are opposite in \eqref{RG1}.

To get dimensionless expressions, 
we use the cutoff $\Lambda=1/\sqrt{\tau}=\tau^{-1/2}$ at time $\tau$ as 
\begin{align}
 &x_\mu = \tau^{1/2}\,\bar x_\mu,\quad
 \partial_\mu = \tau^{-1/2}\,\bar\partial_\mu,
 \quad \phi_x = \tau^{-(d-2)/4}\,\bar\phi_{\bar x},
\end{align}
which gives the relation 
\begin{align}
 V_\tau(\phi) = \tau^{-d/2}\,\bar V_\tau(\bar\phi)
 ~~\mbox{with}~~
 \phi = \tau^{-(d-2)/4}\,\bar\phi .
\label{Vtau}
\end{align}
Here we have placed the bar on quantities 
to indicate that they are dimensionless.  
On the other hand, we use the cutoff 
$\Lambda-\delta\Lambda = 1/\sqrt{\tau+\epsilon}
 = (\tau\,e^{\epsilon/\tau})^{-1/2}$ 
at time $\tau + \epsilon$ as 
\begin{align}
 x_\mu = (\tau\,e^{\epsilon/\tau})^{1/2}\,\bar x_\mu,
 \quad \partial_\mu = (\tau\,e^{\epsilon/\tau})^{-1/2}\,\bar\partial_\mu,
 \quad \phi_x = (\tau\,e^{\epsilon/\tau})^{-(d-2)/4}\,\bar\phi_{\bar x},
\end{align}
which leads to the relation 
\begin{align}
 V_{\tau+\epsilon}(\phi) = (\tau\,e^{\epsilon/\tau})^{-d/2}\,
 \bar V_{\tau+\epsilon}\bigl(\bar\phi\bigr)
 ~~\mbox{with}~~
 \phi = (\tau\,e^{\epsilon/\tau})^{-(d-2)/4}\,\bar\phi.
\label{Vtauepsilon}
\end{align}
Substituting \eqref{Vtau} and \eqref{Vtauepsilon} to \eqref{dimful},  
we finally obtain the following equation 
for the dimensionless local potential 
(we remove the bar from the expression for notational simplicity): 
\begin{align}
 \tau\,\partial_\tau V_\tau(\phi)
 &= \frac{d}{2}\,V_\tau(\phi) - \frac{d-2}{4}\,\phi\,V_\tau'(\phi)
 - V_\tau'(\phi)^2 + B_d\,V_\tau''(\phi) - B_d\,V_\tau''(\phi)^2
\nonumber
\\
 &~~~+ V_\tau'(\phi)\,\int_0^\phi d\phi\,V_\tau''(\phi)^2
  - \frac{d}{2}\,B_d
 \quad \Bigl(B_d\equiv \frac{1}{(4\pi)^{d/2}} \Bigr).
\label{AF_LPA}
\end{align}

\begin{figure}[t]
\begin{center}
\includegraphics[width=15cm]{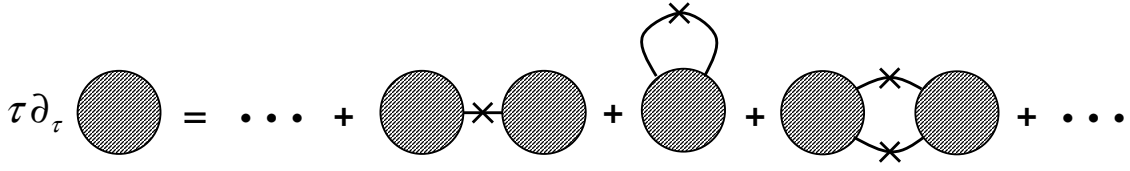}
\begin{quote}
\caption{
A Feynman diagrammatic interpretation of \eqref{AF_LPA}.
The shaded circle represents minus the potential, $-V_\tau(\phi)$.}
\label{feynman}
\end{quote}
\end{center}
\vspace{-6ex}
\end{figure}

Note that the first two terms in \eqref{AF_LPA} 
reflect the simple rescalings of the potential and the field variable. 
The next three terms have a natural interpretation with Feynman diagrams 
(see Fig.~\ref{feynman}). 
In fact, the third term in \eqref{AF_LPA} represents 
the contraction of a propagator in a 1-particle reducible diagram, 
while the fourth term stands for that of 
a propagator in a  1-particle irreducible diagram. 
The fifth term represents the contraction of propagators 
in a 2-particle reducible diagram.

\section{$\varepsilon$ expansion}
\label{sec:epsilon}

The equation \eqref{AF_LPA} can be solved iteratively 
in dimension $d=4-\varepsilon$ with $0<\varepsilon \ll 1$. 
Expanding the potential as 
\begin{align}
 V(\phi) = v_0 + \frac{v_2}{2!}\,\phi^2 + \frac{v_4}{4!}\,\phi^4 + \cdots,
\end{align}
the first few terms in \eqref{AF_LPA} are given by
\begin{align}
 \tau\,\partial_\tau v_2 &= v_2 -2\,v_2^2 + 2\,v_2^3 + B_d\,v_4 -2\,B_d\,v_2 v_4,
\label{v2}
\\
 \tau\,\partial_\tau v_4 &= \frac{\varepsilon}{2}\,v_4 - 8\,v_2 v_4 + 12\,v_2^2 v_4
 -6\,B_d v_4^2 - 2\,B_d\,v_2 v_6 + B_d v_6,
\label{v4}
\\
 \tau\,\partial_\tau v_6 &= (-1+\varepsilon)\,v_6 - 20\,v_4^2 + 76\,v_2 v_4^2
 -12\,v_2 v_6 + 18\,v_2^2 v_6
\nonumber
\\
 &~~~ -30\,B_d\,v_4 v_6
 + B_d\,v_8 - 2\,B_d\,v_2 v_8,
\label{v6}
\\
 \tau\,\partial_\tau v_8 &= \Bigl(-2+\frac{3\varepsilon}{2}\Bigr)\,v_8
 -16\,v_2 v_8 - 112\,v_4 v_6 + 24\,v_2^2 v_8 + 336\,v_4^3
\nonumber
\\
 &~~~ + 464\,v_2 v_4 v_6
 - 56\,B_d v_4 v_8 - 70\,B_d v_6^2.
\label{v8}
\end{align}
In addition to the Gaussian fixed point ($v^\ast_n = 0$), 
a nontrivial fixed point $v_n^\ast$ can be found with the ansatz 
$v_2^\ast=O(\varepsilon)$, $v_4^\ast=O(\varepsilon)$, 
$v_6^\ast=O(\varepsilon^2)$ and $v_n^\ast=O(\varepsilon^3)$ $(n\geq 8)$:
\begin{align}
 v_2^\ast = -\,\frac{1}{36}\,\varepsilon + O(\varepsilon^2),\quad
 v_4^\ast = \frac{1}{36 B_4}\,\varepsilon + O(\varepsilon^2),\quad
 v_6^\ast = -\,\frac{20}{(36 B_4)^2}\,\varepsilon^2 + O(\varepsilon^3),\quad
 v_8^\ast = O(\varepsilon^3).
\end{align}
By linearizing \eqref{v2}--\eqref{v8} around these values, 
the first two eigenvalues are found to be 
$1-\varepsilon/6+O(\varepsilon^2)$ and $-\varepsilon/2+O(\varepsilon^2)$, 
which agree with those of the linearized RG transformation 
at the Wilson-Fisher fixed point 
(note that $-\Lambda\,\partial_\Lambda = 2\,\tau\,\partial_\tau$). 

\section{Conclusion and outlook}
\label{sec:conclusion}

In this paper, we investigated the RG structure 
of the gradient flow. 
To generate the flow, 
instead of using the original bare action,  
we proposed to use the action $S_\tau[\phi]$ at flow time $\tau$. 
We wrote down the basic equation that determines 
the evolution of the action 
and considered a LPA to our equation, 
and showed that the result has a nice interpretation with Feynman diagrams. 
We also made an $\varepsilon$ expansion of the LPA 
and showed that it reproduces the eigenvalues 
of the linearized RG transformation 
around both the Gaussian and the Wilson-Fisher fixed points 
to the order of $\epsilon$.

In order to simplify the argument, 
we have not seriously taken into account the anomalous dimension, 
which actually could be neglected to the order of approximation 
we made in the $\varepsilon$ expansion.  
A careful treatment of the anomalous dimension 
will be given in our forthcoming paper. 
In addition to higher-order calculations of $\varepsilon$ expansion, 
it should be interesting to investigate the LPA of the $O(N)$ vector model.

It is tempting to regard our equation \eqref{flow2b} 
as a sort of exact renormalization group
\cite{Wilson:1973jj,Wegner:1972ih,Polchinski:1983gv,Wetterich:1992yh}
(see \cite{Bagnuls:2001pr,Polonyi:2001se,Rosten:2010vm} 
for a nice review on this subject). 
However, one must be careful in establishing this relationship, 
because the RG interpretation of \eqref{flow2b} is possible 
only when we make a field-variable transformation at every step 
such that the kinetic term is kept in the canonical form 
[see discussions below \eqref{LPA0}]. 
It thus should be interesting to write down an equation 
which incorporates the effect of the change of variable 
in a form of differential equation.

In developing the present work further, 
it must be important to investigate whether the gradient flow  
of the present paper [eq.~\eqref{flow1}]  
also has a nice property in the renormalization of the flowed fields 
and their composite operators. 
In fact, a prominent feature of the conventional gradient flow \eqref{flow0} 
is, as was mentioned in Introduction, 
that there appear no extra divergences in the $(d+1)$-dimensional bulk theory. 
For example, let us consider the expectation value of an operator 
constructed from the flowed field, 
$\mathcal{O}[\phi_\tau]$:  
\begin{align}
 \bigl\langle \mathcal{O}[\phi_\tau] \bigr\rangle_{S_0}
 \equiv \frac{1}{Z_0}\,\int [d\phi_0]\,
 e^{-S_0[\phi_0]}\,\mathcal{O}[\phi_\tau(\phi_0)],
\label{vev_S0}
\end{align}
where $\phi_\tau(\phi_0)$ is the solution to \eqref{flow0}. 
This gives a finite quantity once a proper regularization 
is implemented at the initial cutoff $\Lambda_0$,  
and this absence of extra divergences is attributed to the fact that 
$\phi_\tau(x;\phi_0)$ takes the form  
$\phi_\tau(x;\phi_0)=\int_y\,K_\tau{(x-y)}\,\phi_0(y)+\cdots$.  
Now let us consider the expectation value 
of the same operator $\mathcal{O}[\phi]$ 
with respect to our effective action $S_\tau[\phi]$:
\begin{align}
 \bigl\langle \mathcal{O}[\phi] \bigr\rangle_{S_\tau}
 &\equiv \frac{1}{Z_\tau}\,\int [d\phi]\,
 e^{-S_\tau[\phi]}\,\mathcal{O}[\phi]
\nonumber
\\
 &= \frac{1}{Z_\tau}\,\int [d\phi][d\phi_0]\,e^{-S_0[\phi_0]}\,
 \delta[\phi-\phi_\tau(\phi_0)]\,\mathcal{O}(\phi)
\nonumber
\\
 &= \frac{1}{Z_\tau}\,\int [d\phi_0]\,
 e^{-S_0[\phi_0]}\,\mathcal{O}[\phi_\tau(\phi_0)],
\label{vev_Stau}
\end{align}
where $\phi_\tau(x;\phi_0)$ is now 
the solution to our flow equation \eqref{flow1}. 
Note that this solution also has the form 
$\phi_\tau(x;\phi_0)=\int_y\,K_\tau{(x-y)}\,\phi_0(y)+\cdots$ 
because we make a field-variable transformation at every step 
such that $S_\tau[\phi]$ takes the canonical form, 
$S_\tau[\phi] = \int_x\,\bigl[(1/2)\,(\partial_\mu\phi_(x))^2+\cdots\bigr]$. 
We thus expect that 
two expectation values \eqref{vev_S0} and \eqref{vev_Stau} 
share the same properties 
for the finiteness at short distances.  
We leave the confirmation of this expectation for future work.

Although the present paper only discusses scalar field theory, 
the extension to other field theories should be straightforward. 
The generalization to field theories in curved spacetime 
must also be interesting.

A study along these lines is now in progress 
and will be reported elsewhere. 

\section*{Acknowledgments}
The authors thank 
Daisuke Kadoh, Yoshio Kikukawa, Nobuyuki Matsumoto, Tetsuya Onogi, 
Hidenori Sonoda and Hiroshi Suzuki for useful discussions. 
This work was partially supported by JSPS KAKENHI 
(Grant Number 16K05321).
%
\baselineskip=0.9\normalbaselineskip



\end{document}